\documentstyle[12pt]{article}
\begin{document}
\thispagestyle{empty}
\begin{center}
{\large
The State Research Center of Russian Federation\\
Budker Institute of Nuclear Physics SB RAS\\[20mm]
V.F. Dmitriev and V.B. Telitsin\\[20mm]
MANY BODY CORRECTIONS TO NUCLEAR\\
ANAPOLE MOMENT\\[20mm]
Budker INP 95-107
\vfill
Novosibirsk\\
1995}
\end{center}

\eject
\thispagestyle{empty}
\begin{center}
{\bf Many Body Corrections to Nuclear Anapole Moment }\\[12mm]
{\it Vladimir F. Dmitriev and Vladimir B. Telitsin}\\[10mm]
Budker Institute of Nuclear Physics, \\
630090, Novosibirsk, Russia \\[10mm]
\end{center}

\begin{abstract}
The many body contributions to the nuclear anapole moment of $^{133}$Cs,
$^{205}$Tl, $^{207,209}$PB, and $^{209}$Bi from the core
polarization are calculated in the random-phase approximation with the
effective residual interaction. Strong reduction of a valence nucleon
contribution was found provided by the core polarization effects.
The contribution of the core particles to the anapole moment compensates this
reduction to large extent keeping the magnitude of nuclear anapole moment
close to its initial single particle value.
\end{abstract}

\newpage
\vspace{55mm}
\section{Introduction}
The atomic
parity non-conservation (PNC) effects dependent on nuclear spin are
expected to be dominated by contact electromagnetic interaction of
electrons with nuclear anapole moment (AM) \cite{fk,fks}. The anapole is
a new electro-magnetic moment arising in a system without center of
inversion \cite{zel}. It exists even in such a common object as a
chiral molecule in a state with nonvanishing angular momentum
\cite{kp}. The nuclear anapole moment is induced by PNC nuclear forces.

In all calculations of the anapole moment \cite{fks,dkt,hhm,bp,bp1}
an independent particle model has been used for the nucleus. In this approach,
the AM is determined by the contribution of a single valence nucleon, proton or
neutron. The only attempt to account for configuration mixing has been made
in \cite{bp,bp1}. However, as we shall show below, this is only the part of the
many-body corrections, and is not the dominant one. The contribution arising
from the
induced PNC interaction in the nuclear core is considerably larger. It was not
discussed yet at all. We present here the first treatment of these
kind of effects.

The major coherent effect induced by the residual interaction is the
polarization of the nuclear core by a valence nucleon. This is the main effect
causing the deviation of nuclear magnetic moments from the Schmidt values.
The magnitude of polarization effects depends on the number of transitions from
the core states over the Fermi surface. This number is determined by the
selection rules, i.e. by the tensor rank of the operator.
In the case of the AM, the number of transitions contributing to the core
polarization is greater compared to the magnetic moment, and therefore, larger
renormalization of the AM is expected. Just to illustrate the above statement,
we refer to
renormalization of the M3-octupole moment compared to M1.
In the case of an octupole moment, the number of transitions over the Fermi
surface is much greater, and the core polarization reduces the
valence nucleon contribution to M3 by a factor $\approx$ 4 \cite{dt}

A convenient way to describe the core polarization is to use the effective
renormalized operators, or effective fields, in the terminology of the theory
of
finite Fermi systems \cite{mig}. In the random phase approximation (RPA), the
effective
fields are the solutions of a system of integral equations describing
particle-hole renormalization of the bare vertex. The weak nucleon-nucleon
forces
modify these equations. The modifications effectively produce an additional
contribution to the AM compensating for the strong reduction of the single
particle contribution.

This paper is organized as follows. In the next two sections, we remind the
set of operators contributing to the AM, and discuss the accuracy of the
leading approximation. Later on, we formulate the basic equations, and
introduce the modifications of the equations by the weak nucleon-nucleon
interaction in the leading approximation. Next, we calculate the
renormalization of the single particle contribution, both analytically and
numerically. Finally, we calculate all
contributions to the AM both analytically and numerically, and discuss the
stability of the results under
variation of the constants of the strong residual nucleon-nucleon interaction.

\section{The anapole moment operator}

The anapole moment operator is defined by \cite{fk,fks,khr}
\begin{equation}\label{e1}
{\bf a} = -\pi \int d^3r\ r^2 {\bf j}({\bf r}),
\end{equation}
where ${\bf j}({\bf r})$ is the electromagnetic current density
operator.

The main contribution to the AM comes from the spin part of the current
density. Nevertheless, the other contributions are noticeable and,
apart from magnetization current, we shall discuss below the
contributions from the convection, spin-orbit, and contact currents.
Let us define the corresponding parts of the AM in the following way
\cite{dkt}:
\begin{eqnarray} \label{e2} \nonumber
{\bf a}_s^a & = & \frac{\pi e \mu_a}{m}{\bf r}\times
\mbox{\boldmath $\sigma$} \\
{\bf a}_{conv}^p & = & -\frac{\pi e}{m}\left\{{\bf p},r^2\right\};\;\;
{\bf a}_{conv}^n =0 \nonumber \\
{\bf a}_{ls}^p & = & -\pi
eU_{ls}^{pn}\rho_0\frac{N}{A}r^2\frac{df(r)}{dr}\mbox{\boldmath $\sigma$}\times
{\bf n}
\nonumber \\
{\bf a}_{ls}^n & = & \pi
eU_{ls}^{np}\rho_0\frac{Z}{A}\frac{d(r^2f(r))}{dr}\mbox{\boldmath $\sigma$}
\times{\bf n}
\nonumber \\
{\bf a}^p_c & = & \frac{G}{\sqrt{2}}\frac{\pi e}{m}\rho_0
g_{pn}\frac{N}{A} r^2f(r)\mbox{\boldmath $\sigma$} \nonumber \\
{\bf a}^n_c & = &
\frac{G}{\sqrt{2}}\frac{\pi e}{m}\rho_0 g_{np}\frac{Z}{A} r^2f(r)
\mbox{\boldmath $\sigma$}
\end{eqnarray}
Here $\{\;,\;\}$ is an anticommutator, $\rho_0$ is the central nuclear
density, $f(r)=\rho(r)/\rho_0$ is the nuclear density profile, and
$ U_{ls}^{pn} =U_{ls}^{np} = 134 \, MeV \cdot fm^5 $
is the proton-neutron constant of the effective spin-orbit
residual interaction \cite{dkt}.  The contact current contribution
arises from velocity dependence of the effective nucleon-nucleon weak
forces, taken in the form \cite{dkt,khr}
$$
F_w  =\frac{G}{\sqrt 2} \frac{1}{4m}
\sum_{a,b}\left(\{(g_{ab}\mbox{\boldmath$\sigma$}_a -
g_{ba}\mbox{\boldmath$\sigma$}_b)\cdot({\bf p}_a - {\bf p}_b),
\delta({\bf r}_a -
{\bf r}_b)\}\right.
$$
\begin{equation}\label{e3}
\left. +\, g^{\prime}_{ab}[\mbox{\boldmath$\sigma$}_a\times
\mbox{\boldmath $\sigma$}_b]
\cdot\mbox{\boldmath$\nabla$}\delta({\bf r}_a - {\bf r}_b)\right)
= \frac{1}{2}\sum_{ab} F_w(ab).
\end{equation}
 Interaction (\ref{e3}) generates a mean field weak potential
\begin{equation}\label{e4}
W_a=\frac{G}{\sqrt{2}}\;\frac{g_a\rho_0}{2m}\;
\{ \mbox{\boldmath$\sigma$}\cdot
{\bf p}(r)\},
\end{equation}
where $g_a= g_{ap}\frac{Z}{A}+g_{an}\frac{N}{A}$.

The effective interaction constants $g_{ab}, g_{ba}, g^{\prime}_{ab}$ should
be, strictly speaking, found from experiment. On the other hand, they can
be estimated
from the initial finite range PNC-interaction \cite{ddh} taking zero range
limit
with the account for short range particle particle repulsion \cite{fks,st}.
These "best values"
estimates leads to $g_n \ll 1$, while the constant $g_p $ is approximately 4.5.
The recent discussion of these constants \cite{dn} give, however, different
set for $g_p$ and $g_n$, with $g_p \sim g_n$. Therefore, we shall keep below
the constants explicitly as free parameters.

\section{Single-particle contribution and \protect\newline leading
approximation}
The leading
approximation for the corrections to the single particle wave functions
will be used in calculations of the core polarization effects. Therefore,
it is worth to discuss the accuracy of this approximation.
Neglecting the spin-orbit potential, and assuming constant nuclear density,
we obtain for the correction to the single particle wave function
\cite{cur} \begin{equation} \label{e5} \delta \psi_0 ({\bf r}) = -\imath
\xi_a (\mbox{\boldmath$\sigma$}\cdot {\bf r}) \psi_a ({\bf r}),
\end{equation} where $$ \xi_a = \frac{G}{\sqrt{2}}g_a\rho_0. $$
Using (\ref{e5}), we obtain for the spin part of the AM
\begin{equation} \label{e6}
{\bf a}_s = \frac{Gg\rho_0}{\sqrt{2}}\frac{2\pi e\mu}{m}(R|r^2|R)
\frac{K{\bf I}}{I(I+1)},
\end{equation}
where $ K=(l-I)(2I+1)$; $R$, $l$ and $I$ being the radial wave
function, the orbital
angular momentum of an outer nucleon and
the nuclear spin. It is convenient to discuss AM in terms
of a dimensionless constant $\kappa$ defined as (see Ref.\cite{fks})
\begin{equation}\label{e6'}
<e{\bf a}> = \frac{G}{\sqrt{2}}\frac{K{\bf I}}{I(I+1)}\kappa.
\end{equation}
For the dimensionless constant $\kappa_s$
we have
\begin{equation} \label{e7}
\kappa_s = \frac{9}{10}g\frac{\alpha \mu}{mr_0}A^{2/3},
\end{equation}
where we put $(R| r^2|R) = \frac{3}{5}r^2_0A^{2/3}$.

The naive expression (\ref{e7}) is in rather good agreement with exact
numerical calculations of the spin part of AM \cite{fks}. Therefore, the
approximation is reasonable for averaging the volume type quantities like
$r^2$. For
surface type quantities the situation is quite different. An
instructive example is the convection current contribution to the AM.
Numerical calculation gives for $^{209}Bi \  \kappa_{conv} = -0.019$
while in the leading approximation (\ref{e5}) we obtain the value that
differs by factor $\approx$ 5. The explanation consists in the surface nature
of the convection current contribution   \cite{dkt}
\begin{equation}\label{e8}
\kappa_{conv} = -\pi g\frac{\alpha \rho_0}{mK}
(\delta R|\,r^2\left(\frac{d}{dr} +\frac{K+2}{r}\right)|R).
\end{equation}
The integrand in the matrix element (\ref{e8}) for the outer nucleon with
large angular momentum is peaked at the nuclear surface and
the difference between exact $\delta R(r)$ and its approximate expression
in leading approximation $rR(r)$ provides considerable changes in the
convection current contribution.
\section{RPA renormalization of the AM}
The AM is a T-odd operator.
 Thus, the effective two particle
interaction involved in AM renormalization must change sign under
T-reversal of one of the two particles
$$
T_a F(ab)T^{-1}_a = T_b F(ab)T^{-1}_b = - F(ab)
$$
The simplest interaction satisfying this condition
 is the same spin-spin interaction that changes nuclear
magnetic moments:
\begin{equation} \label{e9}
F_s(ab) = C\,\left( g_0 +
g'_0 \mbox{\boldmath$\tau$}_a \cdot\mbox{\boldmath$\tau$}_b
\right)\mbox{\boldmath
$\sigma$}_a \cdot\mbox{\boldmath$\sigma$}_b \delta ({\bf r}_a - {\bf r}_b).
\end{equation}
Here  $C$ is the normalization constant that we choose according to
\cite{mig} $ C=300 MeV \cdot fm^3$ and the constants $ g' =
1.01 $ and $ g = 0.63$.

The effective interaction between the valence and core particles changes
the interaction of the valence nucleon, with an external field producing
additional core field. In the RPA this effect is accounted for by introducing a
dressed effective vertex $V$ satisfying the equation \cite{mig}
\begin{equation} \label{e10}
V = V_0([{\bf a}_i]) + F\,A\,V.
\end{equation}
Here $V_0([{\bf a}_i]) $ is one of the bare AM operators
Eq.(\ref{e2}); $A$ is the static polarization loop of a particle-hole
pair.
\begin{equation} \label{e11}
A_{\nu_1\nu_1^\prime;\nu_2\nu_2^\prime } = \int
\frac{d\epsilon}{2\pi\imath}\, G_{\nu_1\nu_2}(\epsilon)
G_{\nu_2^\prime\nu_1^\prime} (\epsilon ),
\end{equation}
 $G_{\nu_1\nu_2}(\epsilon)$ being a single particle nucleon propagator.
In Eq.(\ref{e10}) $F$ is the sum of spin-spin interaction Eq.(\ref{e9})
and the weak effective interaction Eq.(\ref{e3}). The propagator
$G_{\nu_1\nu_2}(\epsilon)$  should be calculated in the total mean field
potential including the weak potential (\ref{e4})

It is, however, more convenient to single out the weak interaction
effects, treating them explicitly in first order perturbation theory.
Let $\delta V$ be a correction to the vertex from the weak forces. For
the unperturbed vertex $V$ and the correction $\delta V$ we have the
equations
\begin{equation} \label{e12}
V = V_0([{\bf a}_i]) + F_s\,A\,V,
\end{equation}
\begin{equation} \label{e13}
\delta V = F_w\,AV + F_s\,\delta AV + F_s\,A\delta V.
\end{equation}
Here $F_w$ is the weak nucleon-nucleon interaction (\ref{e3}).
The AM value is given by
\begin{equation} \label{e14}
a = \langle \delta \psi|V|\psi\rangle + \langle\psi|V|\delta \psi
\rangle  + \langle\psi|\delta V|\psi\rangle.
\end{equation}
In the leading approximation, we have
\begin{equation} \label{e15}
a = \imath\xi\langle\psi|[\mbox{\boldmath$\sigma$}\cdot{\bf r},V]|\psi\rangle +
\langle\psi|\delta V|\psi\rangle.
\end{equation}
The first term represents the single particle contribution renormalized
by the spin-spin interaction, while the second term is an additional
contribution from the core particles. Note that the single-particle
contribution is now the expectation value of a transformed renormalized
anapole operator
$V$. The AM is a T-odd operator of E1 type. The commutator of the AM with
$\imath \mbox{\boldmath$\sigma$}\cdot{\bf r}$ transforms it, as well as
$\delta V$,
into a T-odd M1 type operator that
evidently has nonzero expectation value in a state with spin $I$.  The
renormalization effects from the core polarization are different for these
two types of operators. In the next two sections we shall discuss the
renormalization of the electric type single particle AM operators
and the magnetic type operators induced by PNC effects in the core.

Note, that the contact term contribution produces the magnetic type operator
from the very beginning. Therefore, its renormalization is similar to
that of $\delta V$.

\section{Renormalization of the electric type
 single particle operator}
To solve Eq.(\ref{e12}), it is convenient to separate the
angular dependence, introducing a set of tensor operators of rank $J$
\begin{equation} \label{e16}
T^L_{JM} = \{\mbox{\boldmath$\sigma$} \otimes Y_{Lm}\}_{JM}.
\end{equation}
The spin part (\ref{e2}) of the AM is proportional to $T^1_{1M}$. This
is the only T-odd operator of the rank 1 with negative parity. Therefore,
the dressed vertex $V$ will have the same angular dependence as the
bare vertex
$$
V_s = v_s(r)\,T^1_{1M}.
$$
The core polarization effects the radial dependence only, which for
bare spin vertex is just
$$
v^{a0}_s(r) =N^a_s\, r ,
$$
where $N^a_s$ is
$$
N^a_s = \imath \sqrt{\frac{8\pi}{3}}\frac{\pi \alpha \mu_a}{m}.
$$

The dressed vertexes $v_s(r)$ satisfy the following equations:
\begin{equation} \label{e17}
v^a_s(r) = v^{a0}_s(r) + \sum_{b=p,n} g_0^{ab} \int_0^{\infty}r'^2dr'\;
 A^b(r,r')v_s^b (r').
\end{equation}
Here, the constants $g_0^{pp}$ and $g_0^{pn}$ refer to the proton-proton and
proton-neutron spin-spin interaction (\ref{e9}) $g_0^{pp} = g_0 + g'_0$ and
$g_0^{pn} = g_0 - g'_0 $. The normalization interaction constant $C$ is
included in the radial polarization loop which in our case is
\begin{equation} \label{e18}
A(r,r') = \frac{2}{3}C \sum_{jlnj'l'} k_{jln}\,|\langle
jl||T^1_1||j'l'\rangle|^2
R_{jln}(r)R_{jln}(r')G_{j'l'}(r,r';\epsilon_{jln}).
\end{equation}
Here, $k_{jln}$ are the occupation numbers of filled nuclear states,
$\langle jl||T^1_1||j'l'\rangle$ is the reduced matrix element of the
tensor operator, and $G_{j'l'}(r,r';\epsilon_{jln})$ is the Green function
of the radial Schr\"odinger equation with the angular momentum $j',l'$ taken at
the energy of the occupied level $\epsilon_{jln}$.

Before going over to discussion of the numerical results, let us start from a
simple model estimates of the core polarization in a harmonic oscillator
potential without spin-orbit interaction. In order to understand orders
of magnitude, we shall calculate a polarization loop with the bare spin
anapole vertex. Expanding symbolic Eq.(\ref{e12}), we find for the
first-order term
\begin{equation}\label{e12'}
{\bf a}_s^{(1)}({\bf r}) =
C g_s\sigma^j\sum_{\nu\nu'} \psi^{\dagger}_{\nu}({\bf r})\sigma^j\psi_{\nu'}
({\bf r})\frac{k_{\nu} - k_{\nu'}}{\epsilon_{\nu} - \epsilon_{\nu'}}
\langle \nu'|\frac{\pi e\mu}{m}({\bf r}\times\mbox{\boldmath $\sigma$})
   |\nu\rangle.
\end{equation}
Here, $\psi_{\nu}({\bf r})$ are the single-particle
 wave functions. In the absence of
a spin-orbit potential one can sum over spin variables
$$
{\bf a}_s^{(1)}({\bf r}) = -
C g_s\frac{\pi e\mu}{m}\mbox{\boldmath $\sigma$}\times\sum_{\nu\nu'}
\psi^{\dagger}_{\nu}({\bf r})\psi_{\nu'}
({\bf r})\frac{k_{\nu} - k_{\nu'}}{\epsilon_{\nu} - \epsilon_{\nu'}}
\langle \nu'|{\bf r}|\nu\rangle.
$$
 In a harmonic oscillator we have a relation
\begin{equation}\label{e12''}
{\bf r} = - \frac{\imath}{m\omega^2}[H,{\bf p}].
\end{equation}
Using it, we obtain
$$
{\bf a}_s^{(1)}({\bf r}) = -\frac{\imath Cg_s}{m\omega^2}
\frac{\pi e\mu}{m}\mbox{\boldmath $\sigma$}\times\sum_{\nu\nu'}
\psi^{\dagger}_{\nu}({\bf r})\psi_{\nu'}
({\bf r})(k_{\nu} - k_{\nu'}) \langle \nu'|{\bf p}|\nu\rangle
$$
\begin{equation}\label{e13'}
= -\frac{\imath Cg_s}{m\omega^2}
 \frac{\pi e\mu}{m}\mbox{\boldmath $\sigma$}\times [\rho({\bf r}),{\bf p}]=
\frac{Cg_s}{m\omega^2} \frac{\pi e\mu}{m}\mbox{\boldmath $\sigma$}
\times \mbox{\boldmath $\nabla$} \rho({\bf r}).
\end{equation}
Taking the expectation value of the correction in the state with total
angular momentum ${\bf I}$,
we find for the ratio of this correction to the zero-order term
$$
\frac{<\delta \psi |{\bf a}_s^{(1)}|\psi >}{<\delta \psi |{\bf a}_s|\psi >}
=\frac{<\psi |(\mbox{\boldmath $\sigma$}\cdot {\bf r}){\bf a}_s^{(1)}|\psi >}
{<\psi |(\mbox{\boldmath $\sigma$}\cdot {\bf r}){\bf a}_s|\psi >}
$$
\begin{equation}\label{e14'}
=\frac{Cg_s\rho_0}{m\omega^2}
\frac{(R|rf^{\prime}(r)|R)}{(R|r^2|R)} \approx -2.
\end{equation}
Since $f^{\prime}(r)$ is negative, the correction is negative and large.
The sign of the correction is defined by the sign of the spin-spin interaction.
The repulsive interaction decreases the single-particle contribution. For more
realistic potentials, accounting for the spin-orbit potential, we can expect
some changes in this ratio, since
the correction is maximal on the nuclear surface, where the spin-orbit
potential is important. Nevertheless, it remains large.

The results of calculations of the renormalized single particle contribution
are
listed in Table I and Table II for proton and neutron levels. Note the
reduction of the single particle contribution approximately by a factor of 2,
in accordance with the above estimates.

The spin-orbit and contact current contributions differ from the previous
case only by the radial dependence of their bare vertex. Therefore, their
renormalization can be done using the same Eq.(\ref{e17}).

\section{Renormalization of the magnetic type operators }
 Let us now come back to Eq.(\ref{e13}) describing the additional contribution
to the AM coming from parity violation effects in the intermediate states
of the
core particles. Eq.(\ref{e13}) is of the same kind as (\ref{e12})
describing the renormalization of the valence nucleon contribution. The
difference
is in the driving force or the bare vertex, which is no longer connected
to the bare AM operators (\ref{e2}). In calculation of the driving
force, we shall use the leading approximation (\ref{e5}).

Expanding the symbolic notation, we obtain
$$
F_s(ab)\delta A_b V_b = \delta\sum_{\nu\nu'}\int d^3r_b \psi^{\dagger}_{\nu}
({\bf r}_b)F_s(ab)\psi_{\nu'}({\bf r}_b) \frac{k_{\nu}-k_{\nu'}}{\epsilon_{\nu}
-\epsilon_{\nu'}}\langle\nu'|V|\nu\rangle
$$
$$
= \imath\xi_b\sum_{\nu\nu'}\int d^3r_b\left\{ \psi^{\dagger}_{\nu}({\bf r}_b)
 [\mbox{\boldmath$\sigma$}_b\cdot{\bf r}_b,F_s(ab)]\psi_{\nu'}({\bf r}_b)
\langle\nu'|V|\nu\rangle  \right.
$$
$$
\left.+\, \psi^{\dagger}_{\nu}({\bf r}_b)F_s(ab)\psi_{\nu'}({\bf r}_b)
\langle\nu'|[\mbox{\boldmath$\sigma$}\cdot{\bf r},V]|\nu\rangle\right\}
\frac{k_{\nu}-k_{\nu'}}{\epsilon_{\nu}-\epsilon_{\nu'}}
$$
\begin{equation}\label{e19}
=\imath\xi_b\left\{[\mbox{\boldmath$\sigma$}_b\cdot{\bf r}_b,F_s(ab)]\,A_bV_b +
F_s(ab)A_b[\mbox{\boldmath$\sigma$}_b\cdot{\bf r}_b,V_b]\right\}.
\end{equation}
As we see, the driving force in Eq.(\ref{e19}) consists of two different
parts. The first term can be combined with the weak interaction contribution
$F_w(ab)A_bV_b$, giving an effective interaction is the sum of the direct
and inuced weak interactions
\begin{equation} \label{e20}
 {\rm F}_w(ab) = F_w(ab) + \imath\xi_b\,[\mbox{\boldmath$\sigma$}_b\cdot
{\bf r}_b,F_s(ab)].
\end{equation}
The induced weak interaction was first introduced in \cite{fv}.
It has the form
\begin{equation}\label{e21}
F_w^{ind}(ab) = \imath[\xi_a\mbox{\boldmath$\sigma$}_a\cdot{\bf r}_a +
\xi_b\mbox{\boldmath$\sigma$}_b\cdot{\bf r}_b,F_s(ab)],
\end{equation}
and eventually appears in first order calculations of the residual iteraction.
In our case, however, half of interaction (\ref{e21}) enters
equation (\ref{e19}). Using the full interaction (\ref{e21}) will produce
double
counting, because the part related to the valence nucleon is already accounted
for in the $\delta \psi$ correction to the valence nucleon wave function.
The matrix elements of the induced weak interaction are
proportional to the nuclear radius. Therefore, they are enhanced  compared
to the matrix elements of the direct weak interaction by the factor $A^{1/3}$.
For heavy nuclei, this is a considerable factor and, for this reason, we shall
omit below the contribution of the direct term.

The second term in (\ref{e19}) produces  contributions of a different type
to the AM. The commutator $[\mbox{\boldmath$\sigma$}\cdot{\bf r},V]$ produces
the M1 type of
operator, and we can expect its renormalization to be close to that
of the magnetic moment. This very contribution has  been previously
discussed in \cite{bp,bp1}.

According to (\ref{e19}), the correction $\delta V$ can be presented as a
sum of two terms satisfying (\ref{e19}), but with different driving
forces
$$
\delta V = \delta V^{(1)} + \delta V^{(2)},
$$
$$
\delta V^{(1)}_a = \imath\xi_b [\mbox{\boldmath$\sigma$}_b\cdot{\bf r}_b,
F_s(ab)]A_bV_b + F_s(ab)A_b\delta V^{(1)}_b
$$
\begin{equation}\label{e22}
\delta V^{(2)}_a = \imath\xi_b F_s(ab)A_b[\mbox{\boldmath$\sigma$}_b\cdot
{\bf r}_b,V_b] + F_s(ab)A_b\delta V^{(2)}_b.
\end{equation}
It is convenient to use, instead of $\delta V^{(2)}$, another variable related
to it via
\begin{equation}\label{e23}
\chi_a = \imath\xi_a [\mbox{\boldmath$\sigma$}_a\cdot{\bf r}_a,V_a]+\delta
V^{(2)}_a.
\end{equation}
This variable satisfies the equation
\begin{equation}\label{e24}
\chi_a = \imath\xi_a[\mbox{\boldmath$\sigma$}_a\cdot{\bf r}_a,V_a]
+F_s(ab)A_b\chi_b.
\end{equation}
With this definition, we obtain from Eq.(\ref{e15}) the following value
for the anapole moment
\begin{equation}\label{e25}
a =  \langle\psi|\chi + \delta V^{(1)}|\psi\rangle.
\end{equation}
Thus, the contributions to the AM can be presented as the expectation value of
the sum of
two magnetic type operators induced by different driving forces. The first
term represents the contribution of the renormalized magnetic operator obtained
via Michel transformation \cite{cur}, while the second is the
contribution of the induced weak interaction, similar to that of \cite{fv}.

Let us now make an analytical estimate of this correction in the
model used above for the estimates of the renormalization  of single particle
AM operator.
We shall calculate the driving term in Eq.(\ref{e22}) for
$\delta V^{(1)}_a$, using instead of renormalized vertex $V$ the bare spin
vertex defined by Eq.(\ref{e2}). The correction to a proton contribution to the
AM can be presented in the following form
$$
\delta a_s^k \sim \imath\xi_p g_s^{pp}\mu_p\sigma_p^i\sum_{\nu\nu'}
\psi^{\dagger}_{\nu}({\bf r})[\mbox{\boldmath$\sigma$}_p\cdot{\bf r},
\sigma_p^i]\psi_{\nu'}
({\bf r})\langle \nu'|({\bf r}\times\mbox{\boldmath$\sigma$}_p^k)|\nu\rangle
\frac{k^p_{\nu}- k^p_{\nu'}}{\epsilon_{\nu} - \epsilon_{\nu'}}
$$
\begin{equation}\label{e26}
+\imath\xi_n g_s^{pn}\mu_n\sigma_p^i\sum_{\nu\nu'}
\psi^{\dagger}_{\nu}({\bf r})[\mbox{\boldmath$\sigma$}_n\cdot{\bf r},
\sigma_n^i]\psi_{\nu'}
({\bf r})\langle \nu'|({\bf r}\times\mbox{\boldmath$\sigma$}_n)^k|\nu\rangle
\frac{k^n_{\nu}- k^n_{\nu'}}{\epsilon_{\nu} - \epsilon_{\nu'}}.
\end{equation}
In Eq.(\ref{e26}), we have omitted factors common for protons
and neutrons. Calculating the spin commutators we obtain
\begin{equation}\label{e27}
\delta a_s^k \sim \xi_p g_s^{pp}\mu_p\sigma_p^i A_p^{ik}({\bf r}) +
\xi_n g_s^{pn}\mu_n\sigma_p^i A_n^{ik}({\bf r}),
\end{equation}
where
\begin{equation}\label{e28}
A^{ik}({\bf r}) = \sum_{\nu\nu'}\psi^{\dagger}_{\nu}({\bf r})({\bf r}
\times\mbox{\boldmath$\sigma$})^i\psi_{\nu'}({\bf r})\langle\nu'|(
{\bf r}\times \mbox{\boldmath $\sigma$})^k
|\nu\rangle\frac{k_{\nu}-k_{\nu'}}{\epsilon_{\nu} - \epsilon_{\nu'}}.
\end{equation}
Summing over the spin variables, we
obtain for $A^{ik}({\bf r})$
$$
A^{ik}({\bf r}) = \delta_{ik}B^{ll}({\bf r}) - B^{ik}({\bf r}),
$$
where
\begin{equation}\label{e29}
B^{ik}({\bf r}) = \sum_{\nu\nu'}\psi^{\dagger}_{\nu}({\bf r})r^i
\psi_{\nu'}({\bf r})\langle\nu'|r^k
|\nu\rangle\frac{k_{\nu}-k_{\nu'}}{\epsilon_{\nu} - \epsilon_{\nu'}}.
\end{equation}
Using again relation (\ref{e12''}),
we obtain
$$
B^{ik}({\bf r}) = \frac{\imath}{m\omega^2}\sum_{\nu\nu'}\psi^{\dagger}_{\nu}
({\bf r})r^i\psi_{\nu'}({\bf r})\langle\nu'|p^k
|\nu\rangle(k_{\nu}-k_{\nu'})
$$
\begin{equation}\label{e30}
=\frac{\imath}{m\omega^2}\sum_{\nu}
\psi^{\dagger}_{\nu}({\bf r})[r^i,p^k]\psi_{\nu}({\bf r})k_{\nu}
 =-\frac{1}{m\omega^2}\delta_{ik}\rho ({\bf r}),
\end{equation}
where $\rho ({\bf r})$ is the proton or neutron density. Restoring the omitted
 factors, we have for the correction to the AM of a valence proton
\begin{equation}\label{31}
\delta {\bf a}_s({\bf r}) = -\frac{2\pi eC}{m^2\omega^2}
\,(g_s^{pp}\mu_p\xi_p\rho_p({\bf r})+g_s^{pn}\mu_n\xi_n\rho_n({\bf r}))
\,\mbox{\boldmath$\sigma$}
\end{equation}
Its expectation value in a state with nuclear spin $\bf I$ is
\begin{equation}\label{e32}
\delta {\bf a}_s = \frac{2\pi eC}{m^2\omega^2}
\,(g_s^{pp}\mu_p\xi_p\rho_p +
g_s^{pn}\mu_n\xi_n\rho_n)\,\frac{K-\frac{1}{2}}{I(I+1)}{\bf I},
\end{equation}
where $K$ was defined in Eq.(\ref{e6}).
For the ``best values'' of the weak interaction constants \cite{ddh},
$\xi_n \ll \xi_p$ for heavy nuclei. Neglecting $\xi_n$, we obtain for the ratio
of $\delta a_s/ a_s$ for a valence proton
\begin{equation}\label{e33}
\frac{\delta a_s}{a_s} = \frac{2Cg_s^{pp}\rho_0}{m\omega^2
(R|r^2|R)}\,\frac{Z}{A}\,\left ( 1-\frac{1}{2K}\right ),
\end{equation}
where we use the total nuclear density $\rho_0$ instead of the proton
density. Using for $\omega$ the standard value $\omega = 41/A^{1/3}$MeV,
we find
\begin{equation}\label{e34}
\frac{\delta a_s}{a_s} \approx 4.6\,\frac{Z}{A}.
\end{equation}
This result is quite instructive. The contribution to the AM from the induced
weak
interaction is greater than that coming from the single particle weak
potential. However, one should keep in mind that the above calculation has been
performed for the bare anapole vertex.

\section{Results}
The complete results of calculations are summarized in Table I and Table II.
In the first column, we list the results of previous AM calculations in the
independent particle model.   In the second column, we list
the contribution of the valence nucleon renormalized by the strong spin-spin
interaction. On average, the core polarization reduces the single particle
contribution by a factor $\approx$ 2. In the third column, the AM induced by
the core nucleons is written. It significantly restores the reduction of the
single particle contribution, leaving the overall renormalization within
 $\approx$ 10\%.

The spin-spin residual interaction depends on the two constants $g_0$ and
 $g_0'$
corresponding to the interaction in isoscalar and isovector channels.
The major part of the anapole moment is proportional to the nucleon magnetic
moments;
therefore, the isovector part of the AM dominates. For this reason, we can
expect
small sensitivity of the AM to the isoscalar constant $g_0$.
For $^{205}$Tl changing $g_0$ in the interval $0.2 \leq g_0 \leq 0.8$
we get for the AM $0.313 \leq \kappa_{tot} \leq 0.380$. The sensitivity
to the isovector constant $g_0'$ is larger, but still within reasonable
limits. Changing $g_0'$  in the interval $0.5 \leq g_0' \leq 1.5$ we find for
$^{205}$Tl  $0.450 \geq \kappa_{tot} \geq 0.327$.  However, since
the constant $g_0'$ is fixed much better from the fit of
magnetic moment, the above interval for $\kappa_{tot}$ is in fact too large.

To summarize, we have calculated the many body contributions to the nuclear
anapole moment in the random phase approximation with effective nuclear forces.
We found that the contribution to the AM
from the valence nucleon is reduced by a factor $\approx 2$ from the core
polarization
effects. However, PNC effects in the core states produce an additional
contribution
to the AM
that partially compensates the reduction of the single particle AM.
The resulting value of the AM appears to be close to its initial
single-particle value calculated in the independent particle model.

\section{Acknowledgments}
We acknowledge discussions with I.B.Khriplovich,  V.V.Flambaum and
O.P.Sushkov. This work was supported by Russian Foundation for Fundamental
Research grant No.95-02-04436-a.

\eject

\begin{table}
\caption{Anapole moment contributions for proton levels}
\begin{tabular}{l|l|r|rrr}
 & & s.p. &$\langle\delta\psi|V|\psi\rangle$ & $\langle\psi|\delta V|\psi
\rangle$ & Total \\ \hline
 &$\kappa_s$ & 0.070$g_w^p$ & 0.039$g_w^p$ & 0.018$g_w^p+$0.009$g_w^n$ & 0.057
$g_w^p+$0.009$g_w^n$ \\
 &$\kappa_{ls}$& -0.017$g_w^p$& -0.0094$g_w^p$&-0.0035$g_w^p-$0.0004$g_w^n$&-
0.013$g_w^p-$0.0004$g_w^n$ \\
 $^{133}$Cs
 &$\kappa_{conv}$& -0.0041$g_w^p$&-0.0062$g_w^p$&0.0037$g_w^p-$0.0001$g_w^n$&-
0.0025$g_w^p-$0.0001$g_w^n$ \\
 &$\kappa_c$& 0.0066$g_w^{pn}$ &&0.0052$g_w^{pn}-$0.0006$g_w^{np}$&0.0052
$g_w^{pn}-$0.0006$g_w^{np}$ \\
 &              &           &          &                     &              \\
 &$\kappa_{tot}$& 0.050$g_w^p$&0.024$g_w^p$&0.018$g_w^p+$0.008$g_w^n$& 0.041
$g_w^p+$0.008$g_w^n$ \\
 && +0.0066$g_w^{pn}$ && +0.0052$g_w^{pn}-$0.0006$g_w^{np}$ & +0.0052$g_w^{pn}
-$0.0006$g_w^{np}$ \\ \hline
 &$\kappa_s$ & 0.108$g_w^p$ & 0.052$g_w^p$ & 0.048$g_w^p+$0.001$g_w^n$ & 0.100
$g_w^p+$0.001$g_w^n$ \\
 &$\kappa_{ls}$&-0.022$g_w^p$&-0.011$g_w^p$&-0.009$g_w^p-$0.0001$g_w^n$ &-
0.020$g_w^p-$0.0001$g_w^n$ \\
 $^{205}$Tl
 &$\kappa_{conv}$&-0.011$g_w^p$&-0.009$g_w^p$& 0.0009$g_w^p+$0.00002$g_w^n$&
-0.0077$g_w^p+$0.00002$g_w^n$ \\
 &$\kappa_c$& 0.0085$g_w^{pn}$ &&0.0064$g_w^{pn}-$0.0006$g_w^{np}$&0.0064
$g_w^{pn}-$0.0006$g_w^{np}$ \\
 &              &           &          &                     &               \\
 &$\kappa_{tot}$& 0.075$g_w^p$&0.032$g_w^p$&0.039$g_w^p-$0.0003$g_w^n$&-0.018
$g_w^p-$0.0003$g_w^n$ \\
 && +0.0085$g_w^{pn}$ && +0.0064$g_w^{pn}-$0.0006$g_w^{np}$ & +0.0064$g_w^{pn}
-$0.0006$g_w^{np}$ \\ \hline
 &$\kappa_s$ & 0.083$g_w^p$ & 0.039$g_w^p$ & 0.032$g_w^p+$0.0045$g_w^n$ &
0.071$g_w^p+$0.0045$g_w^n$ \\
 &$\kappa_{ls}$&-0.024$g_w^p$&-0.011$g_w^p$&-0.007$g_w^p-$0.0003$g_w^n$ &
-0.020$g_w^p-$0.0003$g_w^n$ \\
$^{209}$Bi
 &$\kappa_{conv}$&-0.005$g_w^p$& 0.0009$g_w^p$& 0.003$g_w^p+$0.00004$g_w^n$&
0.004$g_w^p+$0.00004$g_w^n$ \\
 &$\kappa_c$& 0.010$g_w^{pn}$ &&0.006$g_w^{pn}-$0.0004$g_w^{np}$&0.006$g_w^{pn}
-$0.0004$g_w^{np}$ \\
 &              &           &          &                     &              \\
 &$\kappa_{tot}$& 0.055$g_w^p$&0.029$g_w^p$&0.028$g_w^p+$0.004$g_w^n$&0.057
$g_w^p+$0.004$g_w^n$ \\
 && +0.010$g_w^{pn}$ && +0.006$g_w^{pn}-$0.0004$g_w^{np}$ & +0.006$g_w^{pn}-
$0.0004$g_w^{np}$ \\
\end{tabular}
\end{table}
\begin{table}
\caption{Anapole moment contributions for neutron levels}
\begin{tabular}{l|l|r|rrr}
& & s.p. &$\langle\delta\psi|V|\psi\rangle$ & $\langle\psi|\delta V|\psi
\rangle$ & Total \\ \hline
 &$\kappa_s$ &-0.099$g_w^n$ &-0.052$g_w^n$ &-0.0035$g_w^p-$0.028$g_w^n$ &
-0.0035$g_w^p-$0.080$g_w^n$ \\
&$\kappa_{ls}$&0.006$g_w^n$&0.003$g_w^n$&0.0008$g_w^p+$0.002$g_w^n$ &0.0008
$g_w^p+$0.005$g_w^n$ \\
$^{207}$Pb&
$\kappa_{conv}$& &-0.0002$g_w^n$&0.0001$g_w^p-$0.0002$g_w^n$&0.0001$g_w^p-
$0.0004$g_w^n$ \\
 &$\kappa_c$&-0.002$g_w^{np}$ &&0.0001$g_w^{pn}-$0.001$g_w^{np}$&0.0001
$g_w^{pn}-$0.001$g_w^{np}$ \\
 &              &           &          &                     &             \\
 &$\kappa_{tot}$&-0.093$g_w^n$&0.050$g_w^n$&0.003$g_w^p-$0.027$g_w^n$&-0.003
$g_w^p-$0.076$g_w^n$ \\
 && -0.002$g_w^{np}$ && +0.0001$g_w^{pn}-$0.001$g_w^{np}$ & +0.0001$g_w^{pn}
-$0.001$g_w^{np}$ \\
 \hline
&$\kappa_s$ &-0.062$g_w^n$&-0.033$g_w^n$&-0.004$g_w^p-$0.033$g_w^n$&-0.004
$g_w^p-$0.066$g_w^n$ \\
&$\kappa_{ls}$&0.01$g_w^n$&0.004$g_w^n$&0.001$g_w^p+$0.002$g_w^n$&0.001
$g_w^p+$0.006$g_w^n$ \\
$^{209}$Pb&
$\kappa_{conv}$&&-0.0003$g_w^n$&-0.0005$g_w^p-$0.0002$g_w^n$&-0.0005$g_w^p-
$0.0006$g_w^n$  \\
&$\kappa_c$&0.004$g_w^{np}$ &&0.0003$g_w^{pn}-$0.003$g_w^{np}$& 0.0003
$g_w^{pn}-$0.003$g_w^{np}$  \\
 &              &           &          &                     &           \\
&$\kappa_{tot}$&-0.052$g_w^n$&-0.029$g_w^n$&-0.004$g_w^p-$0.031$g_w^n$&-0.004
$g_w^p-$0.060$g_w^n$  \\
&&0.004$g_w^{np}$ &&+0.0003$g_w^{pn}-$0.003$g_w^{np}$&+0.0003$g_w^{pn}-$0.003
$g_w^{np}$  \\
\end{tabular}
\end{table}

\end{document}